\documentclass[12pt]{article}
\usepackage{biblatex}
\usepackage{amssymb}
\usepackage{openwork}
\title{Either a Confidence Interval Covers, or It Doesn't (Or Does It?): A Model-Based View of Ex-Post Coverage Probability}
\author[1]{Scott Lee}
\date{}
\affil[1]{National Center for Emerging and Zoonotic Infectious Diseases, Centers for Disease Control and Prevention}

\begin{document}

\maketitle

\begin{abstract}
In Neyman's original formulation, a $1-\alpha$ confidence interval procedure is justified by its long-run coverage properties, and a single realized interval is often described only by the slogan that it either covers the parameter or it does not. On this view, post-data probability statements about the coverage of an individual interval are taken to be conceptually out of bounds. In this paper, I present two arguments against treating that "either-or'' reading as the only legitimate interpretation of confidence. The first is intuitive, via a set of thought experiments in which the same joint probability model is used to compute both forward-looking and backward-looking probabilities for occurred-but-unobserved events. The second is more formal, recasting the standard confidence-interval construction in terms of infinite sequences of trials and their associated $0/1$ coverage indicators. In that representation, the design-level coverage probability $1-\alpha$ and the degenerate conditional probabilities given the full data appear simply as different conditioning levels within the same model. I argue that a strict behavioristic reading that privileges only the latter is in tension with the very mathematical machinery used to define long-run error rates. I conclude that the familiar ``either it covers or it does not'' slogan is too restrictive, and that frequentist confidence theory permits a broader class of post-data probability statements than is usually acknowledged. 


\end{abstract}

\section{1 Introduction}

\subsection{1.1 Background }

When Jerzy Neyman introduced his theory of confidence intervals (CIs) in 1937 \cite{neyman1937}, he gave practicing statisticians a strong suggestion for how to interpret them: because \(\theta\) is assumed to be a fixed, unknown constant, once a particular interval is generated, the coverage expression \(\mathbb{P}(L(X)\le \theta \le U(X))\) is mathematically fixed, and so we can only say that the interval either did or did not succeed in covering it. The straightforward mathematical justification for this is that all the randomness in the confidence procedure (CP) lives in the data \(X\), and so once we have a particular realization \(X=x_i\), the expression above becomes degenerate in \(\{0, 1\}\). Intuitively, this also makes sense, since if we imagine sampling a particular set of interval bounds an infinite number of times, the probability of success (under that design) will be either \(0\) or \(1\), depending on whether the original interval covered \(\theta\). As a consequence, practical guidelines for probabilistically interpreting CIs typically revolve around their long-run coverage properties, rather than the properties of any single constructed interval \cite{goodman1994use, hoekstra2014robust, sedgwick2014understanding, hawkins2021use}, and attempts to say otherwise are often, though not always \cite{masson2003using}, branded as errors of interpretation \cite{greenland2016statistical} or fallacies in reasoning \cite{morey2016fallacy}, despite the natural inclination to attach \textit{some} kind of probability to realized intervals ex post (i.e., "post-data"). 

The tension between the accepted interpretation of CIs and the supposedly-fallacious one can be recast more generally as a statement about events we know have occurred, but whose outcomes we have not observed \cite{schield1997interpreting}: for frequentists, randomness, and thus probability, lives in the sampling process and not our knowledge of the outcomes, and so once a sample has been drawn, the ex-ante (i.e., "pre-data") probability has collapsed to some value in \(\{0, 1\}\). Although this is mathematically true, it should also give practicing statisticians some pause, because we happily use frequentist methods for statistical inference in exactly this kind of real-world scenario. . Take, for example, the case of medical diagnosis: given that a patient tests positive on, say, a rapid diagnostic test for the influenza virus, what is the probability that she actually has it? If we stay consistent with our interpretation of CIs, we should also say no probability statement may now be made: given her true underlying health state, the patient either does or does not have the flu, and there is no probability left to assign, because all of the randomness in the sampling process has now been exhausted. Following the "either-or" logic here, though, would substantially limit the clinical value of the diagnostic test in guiding care, and it would obviate the effort epidemiologists and statisticians put in to estimating the test's positive predictive value (\(PPV\)) in the first place, neither of which seems particularly desirable.

The interpretive tension also runs along philosophical lines, with frequentists and propensity-theorists generally taking an ontic view of things (i.e., what matters is how randomness plays out in the world, whether we know about it or not) \cite{hajek2002interpretations}, and Bayesians generally taking an epistemic view (e.g., subjectivists identifying probability with personal degree of belief, or credence \cite{de1937prevision, ramsey1926truth, hajek2002interpretations}) The latter have no trouble accommodating occurred-but-unobserved events, but the former run into more difficulty, since the interpretations tend not to deal explicitly with the role of the observer in making probability assignments (see, e.g., \cite{von1981probability}).  Statistical inference seems to require \textit{some} kind of epistemic component, though--estimation with full knowledge is simply calculation--and even Neyman, arguably one of the foremost operationalists, admits the role of the observer in defining his theory of CIs (if \(\theta\) is fixed-but-unknown, the natural follow-up question is, "Unknown by whom?"). More pointedly, if we do not care whether we know a CI covered \(\theta\), why are we trying to estimate \(\theta\) at all? The answer to this question is perhaps more philosophical than statistical, and it is one that has been addressed thoroughly in the literature on the philosophies of both probability and science, so I will not attempt to summarize it here. However, in the sections below, I hope to show that, strictly speaking, the question itself is not one we need to entertain to provide a formal accounting of occurred-but-unobserved events within frequentism proper, and that we can in fact talk quite sensibly about coverage probability ex post, as long as we state clearly what we mean by the term "probability". 

\subsection{1.2 Paper overview and contributions}

In what follows, I deliberately adopt a rather strict reading of Neyman's slogan--in a nutshell, that ex post probabilities of coverage are entirely out of bounds--and treat it as a normative rule (to be fair, though, this is how many, if not most, instructive pieces handle the interpretation, see e.g., \cite{carlin2008bayesian, piegorsch2015statistical, starnes2010practice, goodman1994use, alshihayb2025some, laven2025common} for examples). The arguments will form a kind of reductio: if we really insist on this kind of rule, we run into uncomfortable constraints on other frequentist uses of probability. The alternative I suggest, and perhaps this paper's main contribution, is that we keep Neyman's (very useful) conception of long-run error control, but that we loosen what we are allowed to say about individual coverage events, especially ex post.

The rest of the paper is structured as follows. In Section 2, I begin the reductio informally by presenting three thought experiments to show some difficulties with inference that arise as a result, in escalating severity from one that makes the underlying probability model not very useful, to one that loses its design-level probabilities entirely. In Section 3, I present a formal argument based on Kolmogorov-style probability theory to show why there is no real mathematical difference between probability statements about events ex ante and ex post, leaning on the machinery of CIs for exemplification. In Section 4, I discuss the implications of the prior sections, presenting a soft normative rule about whether to make intermediate probability assignments ex post and, if so, when. I also suggest a notion of the concept of "confidence" as predictive probability, or as a model-based probabilistic forecast, and I suggest a few directions for future research.

\subsection{1.3 Disclaimers}

\subsubsection{1.3.1 Generative AI usage}
To save time in developing the manuscript, OpenAI's GPT 5.2 Thinking (extended thinking) was used to typeset equations in LaTeX and to conduct a preliminary literature review. To improve readability and ensure correctness, the same model was also used to check the final version of the manuscript for typographical and mathematical errors.

\subsubsection{1.3.2 Data availability statement}
Because no data were used to produce the results in this manuscript, no data has been made available.

\subsubsection{1.3.3 CDC author disclaimer}
The findings and conclusions in this report are those of the author and do not necessarily represent the official position of the Centers for Disease Control and Prevention.

\section{2 Thought Experiments}

\subsection{2.1 Experiment 1: The Physician's Dilemma}

A patient comes into a primary care clinic with a cough, a runny nose, and a fever, all of which she developed within the past day. Because the patient's fever is mild, her physician believes that she has the common cold, but just to be sure, she gives her a rapid antigen test for the flu. Prior clinical testing has shown the rapid test to have an estimated sensitivity of \(0.75\) and a specificity of \(0.98\) relative to polymerase chain reaction (PCR) testing; in this case, PCR has negligible error rates to true disease status, so we may treat it as a direct proxy for the latter. The patient tests positive, and now the physician must decide whether to issue the flu diagnosis and write the patient a prescription for an antiviral, or to tell her she likely has the common cold and recommend a more conservative treatment with over-the-counter medications to reduce congestion, cough, fever, and body aches. 

\textbf{Question}: \textit{What course of action should the doctor recommend?}

Based on the test's known sensitivity and specificity relative to PCR, along with a current estimate of  flu prevalence in the patient's area, the doctor might base her choice on the model-derived probability that the patient has the flu given her positive test result. This figure is known as positive predictive value, or \(PPV\), and can be written generally as \(\mathbb{P}(D=1 \,|\,T=1)\). Assuming a prevalence of \(10\%\), the calculation comes out to
\begin{equation}
\begin{split}
   PPV &= \frac{sens\cdot prev}{(sens\cdot prev) + (1-spec)\cdot(1-prev)} \\
  &= \frac{0.75\cdot0.10}{(0.75\cdot0.10) + 0.02\cdot0.90} \\
  &=0.81
\end{split}
\end{equation}
where \(prev\) is the background prevalence, \(sens\) is the test's sensitivity, and \(spec\) is the test's specificity. Under this model, the patient has a predicted probability of having the flu of  \(81\%\), so the physician strongly considers prescribing her the antiviral and recommending she stay home to recuperate.

Remembering her training in biostatistics, though, the physician also notes that, with this particular patient in her clinic, both a fixed test result \(T_i=t_i\) and a fixed underlying disease state \(D_i=d_i\) have been sampled, and so there is no randomness left to assign the patient a probability of disease--she either has the disease, or she does not. From this point of view, \(PPV\) turns into the degenerate conditional \(\mathbb{P}(D_i=1 \mid T_i=t_i, D_i=d_i)=\mathbf{1}\{d_i=1\}\in\{0,1\}\), and now the physician must base her recommendation on an underlying truth value she has no possible way of knowing. Realizing how strange this would be, she decides to use the model-derived number of \(81\%\) and recommends a course of action that assumes the patient truly has the flu.

\subsection{2.2 Experiment 2: The Cat Tasting Treats}

A cat named Sophie loves treats. To keep her happy, her owner buys her a big box of mixed-flavor treats from the pet supplies store. The label on the box guarantees that \(75\%\) of the treats are seafood-flavored, and that the remaining \(25\%\) of the treats are chicken-flavored. (Apparently, to make this guarantee, the company producing the treats stamps each one with a unique random ID number and then uses a perfectly reliable computer vision system to track which one ends up in which box, which then lets them automatically verify the flavor composition of the boxes.)

From past observation, the owner knows that Sophie prefers the seafood-flavored treats, purring \(80\%\) of the time after eating them but only \(60\%\) of the time after eating the chicken-flavored treats. The purring has a carryover effect to what she does after eating the treat, too: if she purrs, there is a \(90\%\) chance that she will take a nap, and only a \(10\%\) chance that she will roam the house looking for something else to eat. If she does not purr, though, she is much more inclined to keep foraging, and the two chances align at \(50\%\).

The owner draws a treat from the box, takes note of its number (in this case, \(\#123\)), and puts it on the ground.

\textbf{Question A}: \textit{Assuming the owner has no idea what flavor the treat is, what is the probability that Sophie ends up taking a nap?}

As before, we can base our answer on two model-based probabilities: the unconditional, which marginalizes the purring and napping probabilities over the distribution of flavors in the box; or the conditional degenerate, which considers the treat's flavor fixed but unknown. Under the former, the cat's probability of taking a nap is
\begin{equation}
\begin{split}
P(\text{nap})
&= P(F = \text{sea})\,P(\text{nap} \mid F = \text{sea})
 + P(F = \text{chk})\,P(\text{nap} \mid F = \text{chk}) \\
&= 0.75 \,P(\text{nap} \mid F = \text{sea})
   + 0.25 \,P(\text{nap} \mid F = \text{chk}) \\
&= 0.75 \big[ P(\text{purr} \mid F=\text{sea}) P(\text{nap} \mid \text{purr})
             + P(\text{no purr} \mid F=\text{sea}) P(\text{nap} \mid \text{no purr}) \big] \\
&\quad + 0.25 \big[ P(\text{purr} \mid F=\text{chk}) P(\text{nap} \mid \text{purr})
                  + P(\text{no purr} \mid F=\text{chk}) P(\text{nap} \mid \text{no purr}) \big] \\
&= 0.75 \big[ 0.8 \times 0.9 + 0.2 \times 0.5 \big]
  + 0.25 \big[ 0.6 \times 0.9 + 0.4 \times 0.5 \big] \\
&= 0.75 \times 0.82 + 0.25 \times 0.74 \\
&= 0.615 + 0.185 \\
&= 0.80.
\end{split}
\end{equation}
A fairly high probability at \(80\%\), but one that makes sense, given that most treats in the box are the cat's favorite, that she purrs \(80\%\) of the time after eating them, and that she naps \(90\%\) of the time after purring. On the other hand, when basing the calculation off of the conditional degenerate law
\begin{equation}
P(F = \text{sea} \mid F = F_{123})
=
\begin{cases}
1, & F_{123} = \text{sea},\\
0, & F_{123} = \text{chk},
\end{cases}
\in \{0,1\},
\end{equation}
the \(80\%\) probability forks into two, with
\begin{equation}
\begin{split}
P(\text{nap} \mid F_{123} = \text{sea})
&= P(\text{purr} \mid F=\text{sea}) P(\text{nap} \mid \text{purr})
 + P(\text{no purr} \mid F=\text{sea}) P(\text{nap} \mid \text{no purr}) \\
&= 0.8 \times 0.9 + 0.2 \times 0.5 \\
&= 0.82, \\[6pt]
P(\text{nap} \mid F_{123} = \text{chk})
&= P(\text{purr} \mid F=\text{chk}) P(\text{nap} \mid \text{purr})
 + P(\text{no purr} \mid F=\text{chk}) P(\text{nap} \mid \text{no purr}) \\
&= 0.6 \times 0.9 + 0.4 \times 0.5 \\
&= 0.74.
\end{split}
\end{equation}
The treat's flavor is indeed a fact of the world, and the forked probability is mathematically correct. However, since the purring and napping have yet to happen, and since our model gives us a clear way of quantifying the uncertainty in those outcomes with a single number, the unconditional is probably the way to go (I imagine it is also the number that most statisticians would use to answer this question).

\textbf{Question B}: \textit{The owner leaves the room so Sophie can eat in peace, then comes back a few minutes later to find her napping on the couch. What is the probability the treat was seafood-flavored?}

This one is easy, since all we need to do is use Bayes' rule to calculate the required probability:
\begin{equation}
\begin{split}
P(F = \text{sea} \mid \text{nap})
&= \frac{P(\text{nap} \mid F=\text{sea})\,P(F=\text{sea})}
         {P(\text{nap})} \\
&= \frac{0.82 \times 0.75}{0.80} \\
&= \frac{0.615}{0.80}
= \frac{123}{160}
\approx 0.77.
\end{split}
\end{equation}
The motive for considering this conditional is that it is exactly the quantity our model assigns to the treat’s hidden flavor given the final outcome of the trial. For a particular treat, this probability might be described as epistemic, as it represents the owner’s uncertainty about the flavor, not any physical randomness in the world. Note, though, that in the actual sequences of events described by the model, the flavor is fixed before the cat eats, purring (or not) follows, and napping (or roaming) comes last. The same joint probability model governs all of these variables together, and so if we are willing to use that model \textit{before} sampling to compute forward-looking probabilities like \(P(\text{nap})\), by the same mathematical rules, we are equally entitled to use it \textit{after} the fact to compute backward-looking probabilities like \(P(F = \text{sea} \mid \text{nap})\), regardless of whether we view them as epistemic. Rejecting the latter while accepting the former would amount to using only part of the model’s probabilistic structure, even though both have well-defined frequentist properties over repeated sampling.

\subsection{2.3 Experiment 3: Deep Truffle}

An artisanal chocolatier is scaling up operations and buys a new set of machines to help her make more truffles. Her setup contains:

\begin{enumerate}
    \item A fabricator that produces chocolate truffles, which has two components: one that creates the hollow chocolate shell, and one that fills the shell with chocolate ganache. The fabricator always produces a shell, but the filler doesn’t always activate, failing to inject any ganache at all \(10\%\) of the time. The other \(90\%\) of the time it works, though, and when it does, an auto-stop mechanism detects when the shell is full to prevent over-filling.
    \item A second machine that weighs the truffles and detects the ones that are hollow, which are lighter. This machine is more reliable than the filler mechanism in the first, but it’s still not perfect, incorrectly reading filled truffles as hollow \(5\%\) of the time (false positives) and hollow truffles as filled \(1\%\) of the time (false negatives).
    \item A conveyor belt that sends truffles from the first machine to the second.
    \item A basic pressure sensor that detects whether any truffle is on the conveyor belt and emits a small single beep, if so (the sensor is perfectly accurate).
    \item A second conveyor belt that can send truffles from the weigher back to the filler as needed.
\end{enumerate}

If the second machine detects a hollow truffle, it sends an electrical signal to the fabricator instructing it to pause the shell-maker, and it returns the suspect truffle along the second conveyor belt to the fabricator to be filled. Once the suspect truffle arrives, the fabricator attempts to fill the truffle as if it were a fresh shell, with the filler activating with exactly the same timing and probability (\(90\%\)). If the filler activates and the truffle is truly hollow, it always fills it; if the truffle is already full, however, the filler’s sensor detects the fullness and does not inject more ganache. The fabricator then emits the truffle and sends it back down the belt to be weighed, all with exactly the same timing as if the truffle had been created from scratch.

The chocolatier activates the machines, and the pressure sensor beeps shortly thereafter, indicating the fabricator has emitted a truffle. 

\textbf{Question}: \textit{With this specific truffle on the belt, but the weigher having not yet made its measurement, what is the probability that the next truffle the fabricator emits is correctly filled? }

Sticking only to design-level information, we can run the calculation like this:
\begin{equation}
\begin{aligned}
P(\text{next filled})
  &= P(W = \text{filled}) \cdot 0.9
   + P(W = \text{hollow}) \cdot
     P(\text{next filled} \mid W = \text{hollow}) \\[4pt]
  &= \bigl(0.9 \times 0.95 + 0.1 \times 0.01\bigr) \cdot 0.9 \\[2pt]
  &\quad
   + \bigl(0.9 \times 0.05 + 0.1 \times 0.99\bigr) \cdot
     P(\text{next filled} \mid W = \text{hollow}) .
\end{aligned}
\end{equation}

\begin{equation}
\begin{aligned}
P(S = \text{filled} \mid W = \text{hollow})
  &= \frac{0.9 \times 0.05}{0.9 \times 0.05 + 0.1 \times 0.99}
   = \frac{5}{16}, \\[4pt]
P(S = \text{hollow} \mid W = \text{hollow})
  &= \frac{11}{16}.
\end{aligned}
\end{equation}

\begin{equation}
\begin{aligned}
P(\text{next filled} \mid W = \text{hollow})
  &= 1 \cdot P(S = \text{filled} \mid W = \text{hollow}) \\
  &\quad + 0.9 \cdot P(S = \text{hollow} \mid W = \text{hollow}) \\[4pt]
  &= 1 \cdot \frac{5}{16}
   + 0.9 \cdot \frac{11}{16}
   = \frac{149}{160}
   \approx 0.93125.
\end{aligned}
\end{equation}

\begin{equation}
\begin{aligned}
P(\text{next filled})
  &= 0.856 \cdot 0.9
   + 0.144 \cdot \frac{149}{160} \\[4pt]
  &= \frac{1809}{2000}
   \approx 0.9045.
\end{aligned}
\end{equation}

If, however, we decide to condition the probability on the realized fact that the current truffle either is or is not filled, we end up with the same kind of forked probabilities we saw in the cat-treat experiment:
\begin{equation}
\begin{aligned}
P(\text{next filled} \mid S = \text{filled})
  &= P(W = \text{filled} \mid S = \text{filled}) \cdot 0.9 \\
  &\quad
   + P(W = \text{hollow} \mid S = \text{filled}) \cdot 1 \\[4pt]
  &= 0.95 \times 0.9 + 0.05 \times 1
   = 0.905, \\[8pt]
P(\text{next filled} \mid S = \text{hollow})
  &= P(W = \text{filled} \mid S = \text{hollow}) \cdot 0.9 \\
  &\quad
   + P(W = \text{hollow} \mid S = \text{hollow}) \cdot 0.9 \\[4pt]
  &= 0.01 \times 0.9 + 0.99 \times 0.9
   = 0.9.
\end{aligned}
\end{equation}
Unlike in the cat-treat experiment, though, the forking here leads to an apparent puzzle where we now have two probabilities attaching to \emph{exactly} the same underlying event: the next truffle's fill status. By design, every emitted truffle's fill status after the first, including that of the one presently upcoming, is governed ex ante by \(P(\text{next filled}) \approx 0.9045\). At the same time, conditioning on the realized fact that the current truffle either is or is not filled yields the two conditional values \(0.905\) and \(0.9\). The statement ``either the current truffle is now filled, or it is not'' is necessarily true, but it \emph{cannot} be privileged as the only admissible post-trial probability content for this event, because doing so would prohibit the very quantity \(P(\text{next filled})\) that the model instantiates by design.

\subsection{2.4 A quick disclaimer}

The foregoing examples are not meant to describe how working statisticians actually reason about occurred-but-unobserved events in practice--clearly, and especially in the case of $PPV$, they do not. What they are meant to show, however, is that taking Neyman's "either-or" slogan seriously leads us to a very real tension between what we are told to say, what we would like to say, and what our models should allow us to say. In the following section, I take a look at the mathematics underlying this tension and build toward what I hope is a reasonable way of resolving it.

\section{3 Infinite sequences, microstates, and coverage}
\label{sec:microstates}

Below, I briefly recall the mathematical definition of a confidence procedure and then recast it in terms of \emph{microstates}—fully fixed infinite sequences of trials—in order to make explicit how the design-level coverage probability $1-\alpha$ relates to those sequences. The primary goal here is to set the stage for making a principled claim about whether, if ever, the "either-or" statement Neyman makes is compatible with the frequentist notion of probability, and, if so, when.

\subsection{3.1 Confidence intervals and coverage indicators}

To begin, we fix a parametric model
\[
(\Omega,\mathcal{F},\{P_\theta : \theta \in \Theta\}),
\]
with parameter $\theta \in \Theta$ treated as fixed but unknown. Now, let
\[
X : (\Omega,\mathcal{F},P_\theta) \to \mathcal{X}
\]
denote the data, with distribution $P_\theta^X$ under $P_\theta$. Under this setup, a $(1-\alpha)$ confidence interval for $\theta$ is a measurable map
\[
I : \mathcal{X} \to \mathcal{I}, \qquad x \mapsto I(x) = [L(x),U(x)],
\]
such that for every fixed $\theta \in \Theta$,
\begin{equation}
  P_\theta\big( \theta \in I(X) \big)
  \;=\;
  P_\theta\big( L(X) \le \theta \le U(X) \big)
  \;=\;
  1-\alpha.
  \label{eq:coverage-def-micro}
\end{equation}

Since coverage is a yes-no event, we can also define the \emph{coverage indicator}
\begin{equation}
  Z_\theta(X)
  \;:=\;
  \mathbf{1}\{\theta \in I(X)\}
  \;=\;
  \mathbf{1}\big\{L(X) \le \theta \le U(X)\big\},
  \label{eq:Z-def-micro}
\end{equation}
which is a $\{0,1\}$-valued random variable on $(\Omega,\mathcal{F},P_\theta)$. From here, we can see that \eqref{eq:coverage-def-micro} is equivalent to
\begin{equation}
  E_\theta\big[ Z_\theta(X) \big] = 1-\alpha,
  \qquad\text{or}\qquad
  Z_\theta(X) \sim \mathrm{Bernoulli}(1-\alpha)
  \label{eq:Z-bernoulli-micro}
\end{equation}
under $P_\theta$. 

\subsection{3.2 Infinite sequences of experiments}

To return to the concept of microstates, we can imagine embedding the interval-generating procedure in an infinite sequence of experiments. For a fixed $\theta \in \Theta$, let
\[
X_1, X_2, \dots \;\sim\; P_\theta^X
\]
be i.i.d.\ copies of the data, and then define
\[
I_i := I(X_i) = [L(X_i),U(X_i)],
\qquad
Z_i := \mathbf{1}\{\theta \in I_i\}
     = \mathbf{1}\{L(X_i) \le \theta \le U(X_i)\},
\quad i=1,2,\dots.
\]
By \eqref{eq:Z-bernoulli-micro}, each $Z_i$ has the same Bernoulli$(1-\alpha)$ law under $P_\theta$, and $(Z_i)_{i\ge 1}$ are independently and identically distributed (i.i.d.). In particular,
\begin{equation}
  P_\theta(Z_i = 1) = 1-\alpha,
  \qquad i=1,2,\dots.
  \label{eq:design-coverage-each-i-micro}
\end{equation}
Each outcome $\omega \in \Omega$ determines an infinite \emph{microstate path}
\[
\bigl( X_1(\omega), X_2(\omega), \dots \bigr),
\quad
\bigl( I_1(\omega), I_2(\omega), \dots \bigr),
\quad
\bigl( Z_1(\omega), Z_2(\omega), \dots \bigr),
\]
with $Z_i(\omega) \in \{0,1\}$ for every $i$. From this viewpoint, a single realized world $\omega$ contains a fully-fixed infinite sequence of intervals and coverage outcomes; all of the randomness in the model lives in the draw of $\omega$ itself, which determines every element of the sequence simultaneously. The strong law of large numbers applied to $(Z_i)_{i\ge 1}$ yields
\begin{equation}
  \frac{1}{n} \sum_{i=1}^n Z_i
  \;\xrightarrow{\text{a.s.}}\;
  1-\alpha,
  \label{eq:slln-micro}
\end{equation}
so that with $P_\theta$-probability $1$, the realized microstate path has an asymptotic coverage fraction equal to the design-level $1-\alpha$. 

\subsection{3.3 Connection to the Law of Iterated Expectation}

The formalization above helps clarify two related points. First, for the \(i\)th use of the confidence procedure, the model simultaneously implies
\[
P_\theta\bigl(Z_\theta(X_i) = 1\bigr) = 1 - \alpha
\]
and
\[
P_\theta\bigl(Z_\theta(X_i) = 1 \,\big|\, X_i\bigr)
= Z_\theta(X_i) \in \{0,1\} \quad \text{almost surely}.
\]
Equivalently, using a regular conditional version, for \(P_\theta^{X_i}\)-almost every realized sample \(x_i\),
\[
P_\theta\bigl(Z_\theta(X_i) = 1 \,\big|\, X_i = x_i\bigr)
= \mathbf{1}\{\theta \in I(x_i)\} \in \{0,1\}.
\]
Thus the same model supports both the unconditional design-level statement and the degenerate conditional statement given the realized sample. In probabilistic terms, they are related by the law of iterated expectation:
\[
E_\theta\bigl[Z_\theta(X_i)\bigr]
= E_\theta\Bigl\{E_\theta\bigl[Z_\theta(X_i)\mid X_i\bigr]\Bigr\}
= 1 - \alpha.
\]
Reporting \(P_\theta(Z_\theta(X_i) = 1)=1-\alpha\) amounts to working at a coarser information level, whereas reporting \(P_\theta(Z_\theta(X_i) = 1 \mid X_i)\) uses the finer \(\sigma\)-algebra generated by \(X_i\). In this sense, the familiar either-or reading corresponds to privileging the finest conditioning level, not to any failure of the model to support the unconditional design-based statement.

The second thing the formalization helps us see is that, in the infinite-sequence representation, each use of the procedure corresponds to an index $i$ and an interval $I_i$. For every $i$, the model assigns the same design-level coverage
\[
P_\theta\bigl(Z_\theta(X_i) = 1\bigr) = 1 - \alpha,
\]
even though in each realized world $\omega$ the interval $I_i(\omega)$ has fixed endpoints. In the finite-support setting, this observation becomes especially compelling: because the $Z_i$ are independent and each interval realization with positive mass under the design satisfies $P_\theta(X_i = x^{*}) > 0$ for every $i$, the Second Borel--Cantelli lemma guarantees that any such realization $x^{*}$ will recur infinitely often almost surely. That is, if we have already constructed an interval $I_j = I(x^{*})$ with particular numeric endpoints at some earlier index $j$, there will almost surely be a later index $i > j$ at which $X_i = x^{*}$ and $I_i = I(x^{*})$ exactly. At that later index, the model assigns the same design-level coverage probability as it always does:
\[
P_\theta\bigl(Z_\theta(X_i) = 1\bigr) = 1 - \alpha.
\]
But notice what this means: the observer already knows the numeric values of the endpoints that $I_i$ will take, having seen them when $I_j$ was constructed, and yet the model's probability assignment is unchanged. Merely knowing the interval's endpoints---the very information that the "either-or" reading treats as collapsing
coverage to $\{0,1\}$---does not, and cannot, alter the design-level probability that the model attaches to the coverage event. What this suggests is that the familiar post-data collapse to $\{0,1\}$ is not forced on us by anything intrinsic to having observed the data; rather, it reflects a choice to condition on the finest available
$\sigma$-algebra, when the model equally supports the coarser one under which coverage remains $1-\alpha$.

\section{4 Discussion}

\subsection{4.1 Competing ex post interpretations}

Above, I revisited the two kinds of coverage probabilities given to us by the model underlying a confidence procedure: the design-level unconditional, and the degenerate conditional based on a realized interval's endpoints. If both kinds of probabilities are available under the model, do they fare equally well as \textit{interpretations} of coverage ex post? And can the degenerate conditional be the only one we allow, to the exclusion of all others? I would like to suggest the answer to both questions is "no". To see why, it will be helpful to return to the Deep Truffle thought experiment from Section 2.3. 

\subsubsection{4.1.1 The degenerate conditional as one interpretation}

In order to answer the question, "What is the probability the next emitted truffle is filled?", there are two options for us to consider. On the one hand, there are two forked conditional probabilities for the next truffle—one conditional on the current truffle in fact being filled, and one conditional on it in fact being hollow. On the other hand, there is a single design-level probability, obtained by averaging those two conditionals with respect to the model’s probability that the current truffle is filled. Mathematically, both constructions are legitimate. Interpretationally, however, I would argue that they are not on equal footing. Under a frequentist view, probability is defined with respect to a single, typically infinite sequence of experiments conducted under repeatable conditions (i.e., a reference class, or, in von Mises’ terms, a “collective”, a term which I use here because it more clearly captures the notion of an infinite sequence \cite{von1981probability, gillies2012philosophical}). In the chocolatier's situation, once we condition on the truth value of a realized event (“this truffle is full” versus “this truffle is hollow”), a single infinite collective is no longer available to underwrite \textit{all} of the probabilities we might like to assign to future events. To interpret both forked probabilities as long-run frequencies, we must imagine two distinct collectives, each with its own version of the Markov chain above describing the truffle-emission-and-weighing process. These collectives correspond to two incompatible futures—one in which the present truffle is (and always was) filled, and one in which it is hollow--but because the current truffle cannot actually be both filled and hollow, only one of these can coincide with the actual state of affairs. As Hajek points out, we need to choose a single reference class for obtaining the probabilities \cite{hajek2007reference}, and unless we entertain a kind of frequentist inference based on simultaneous collectives, we are at something of an impasse.

An analogous issue arises for confidence intervals. As an example, let us imagine that we would like to calculate the probability that both our current interval and the next interval we construct will both cover \(\theta\). Let $C_i$ denote the indicator that the $i$th interval produced by a fixed procedure covers $\theta$, so that $C_i \sim \mathrm{Bernoulli}(1-\alpha)$ and $(C_i)_{i\ge 1}$ are i.i.d.\ under the model. At the design level we have
\[
\Pr(C_1 = 1, C_2 = 1) = (1-\alpha)^2,
\]
which can be read, in frequentist terms, as the long-run proportion of pairs of consecutive intervals in which both cover $\theta$. Once a particular sample has been observed and a particular interval constructed, however, we can again be tempted to introduce forked, ex post readings of this joint probability. If the present interval in fact covers $\theta$, then in the model
\[
\Pr(C_1 = 1, C_2 = 1 \mid C_1 = 1)
  = \Pr(C_2 = 1 \mid C_1 = 1)
  = 1-\alpha,
\]
where the last equality follows from the independence of $(C_i)$. If instead the present interval in fact fails to cover, then
\[
\Pr(C_1 = 1, C_2 = 1 \mid C_1 = 0) = 0.
\]
Mathematically, these probabilities are all perfectly well defined, but again, under a strict frequentist account, probabilities are typically interpreted with respect to a single infinite sequence of repetitions. For CIs, the unconditional design-level probability $(1-\alpha)^2$ corresponds to that single collective. By contrast, the two forked readings of the joint probability implicitly appeal to two distinct collectives: one in which the realized first interval covers $\theta$ (yielding a joint probability of $1-\alpha$), and one in which it does not (yielding a joint probability of $0$). As with the truffles, only one of these can coincide with the actual state of affairs in the world, since the realized interval is not simultaneously covering and non-covering, and again, unless we are willing to treat the ensuing forked conditionals as living in separate collectives, the unconditional joint probability $(1-\alpha)^2$ is probably the safer bet, understood again at the design level rather than as a statement about the specific realized pair of intervals. (If this seems like a toy example, we might also imagine wondering how many of, say, $1000$ CIs we have just constructed will cover the parameter--without admitting a design-level $p$ to plug in to the equation for the binomial mean, that question becomes a very difficult one to answer with a precise number).

\subsubsection{4.1.2 The degenerate conditional as the only interpretation}

The conditional degenerate is certainly one interpretation we can adopt ex post, but can it be the \textit{only} one? I suggest that the answer here must be "no", and precisely for this reason: if it were the only interpretation we could adopt, then we would then lose the ability to make design-level probabilistic statements about future events conditioned on the current state of the affairs. The thought experiments in Section 2 make this consequence clear, especially in the case of Deep Truffle, where insisting that the degenerate conditional is the only possibility for the current truffle's fill status prevents us at all from recovering the design-level probability that the next truffle will be filled. Allowing both interpretations side-by-side would be perfectly fine, since we would still be able to say that the current truffle is either filled or hollow, if we wished, but we would have the option of backing off to the coarser-grained unconditional probability to estimate the fill status of the next truffle, as our model requires, and as common practice might suggest. The question here is really not whether the Neyman-style interpretation is legitimate (it is) or should be allowed alongside the others our model defines (it should), but rather whether it should be the only one that is allowed, to the exclusion of all others--again, in light of the evidence above, it seems reasonable here to say "no".

\subsection{4.2 Toward an interpretation of confidence}

The discussion above suggests that part of the trouble comes from conflating distinct layers of probability that the model makes available. We clearly have at least two model-based layers to work with when interpreting confidence intervals: at the design level, there is the unconditional coverage probability $1 - \alpha$ attached to the procedure as a whole, as well as each of its constituent runs; and at the ex post level, conditioned on the truth value of the coverage event for a particular interval, there are the degenerate probabilities $0$ and $1$ corresponding to non-coverage and coverage, respectively. Both are properties of the same probability measure $P_\theta$, viewed at different conditioning levels. The design-level quantity arguably holds up better under scrutiny, as it avoids the multi collective complications noted above. For some distributions and estimators, though, a bare report of $1-\alpha$ can feel somewhat strange ex post. The ``trivial'' interval, which returns the real line with probability $1 - \alpha$ and the empty set with probability $\alpha$, is perhaps the most familiar example, but Welch's uniform confidence interval~\cite{welch1939confidence} and Basu's construction based on a $\theta$-free $X$~\cite{dasgupta2010ancillary} are also instructive.

In such cases, I would suggest that, while the two model-based layers are certainly still in play---considered only as a random draw, any interval generated by the procedure does, by definition, inherit the design-level coverage probability---it may be more natural to admit a third layer of probability: that of a single-case predictive probability of the coverage event itself, made in light of the information available. From this perspective, our $1 - \alpha$ report ex ante can be understood as a default model-based assessment of the coverage indicator for the interval to be constructed, distinct from the interval's realized coverage status under the model. In some cases, features of both the design and the numerical value of the realized interval may suggest a more refined ex post assessment. In that richer information state, the relevant probability need not equal $1 - \alpha$, but it also need not collapse to $\{0,1\}$: it may instead be an intermediate probability statement about the coverage event for the realized interval relative to other intervals generated under the design that are similar with respect to some $\theta$-free feature of the data or interval.

Along these lines, I suggest that this kind of predictive assessment is, in essence, what Neyman's usage of the term ``confidence'' was gesturing toward: a non-oracle observer's best guess as to how many intervals like the one we have constructed, or the ones we are about to construct, will succeed in capturing the parameter. Separating our probabilistic views of confidence procedures in this way---into (i) design-level probabilities under $P_\theta$, (ii) degenerate conditionals under $P_\theta$ given the full data, and (iii) information-indexed probability statements about the coverage indicator---would ease some of the tension in interpreting constructed intervals ex post, since we could then specify which view we are referring to when we talk about, for example, the ``probability'' that a single interval has covered the parameter.

To explore this third perspective more fully, including a more explicit mathematical treatment, I take it up in a separate companion paper \cite{lee2026confidence}.

\subsection{4.3 A proposal for ex post probability statements}

Leaving the notion of confidence aside, I would like to end by posing a soft rule for deciding how to answer questions about the probability of occurred–but–unobserved events: only condition on post–trial information when it actually reduces uncertainty about the outcome. In formal terms, we may view this as choosing an information $\sigma$–algebra $\mathcal{G}$ with
\[
\sigma(\text{design}) \;\subseteq\; \mathcal{G} \;\subseteq\; \sigma(\text{full microstate}),
\]
and taking our ex post probabilities to be $P_\theta(\,\cdot \mid \mathcal{G})$, where $\mathcal{G}$ is the information we happen to have at hand, rather than conditioning on the maximal $\sigma$–algebra that renders the outcome event measurable (and hence degenerate). Choosing such a $\mathcal{G}$ keeps our ex post probability statements within the hierarchy of $\sigma$–algebras defined by the model, avoids degeneracy, and preserves the collective originally implied by the design. For a cat treat with unknown flavor, a patient with an unknown disease status, or a constructed CI that gives no clues as to coverage, the only information we have post–trial is simply that sampling was performed according to the experimental procedure—a probability–one event under the model—which therefore leaves the pre–trial probability of success unchanged. As demonstrated above, conditioning instead on the hidden outcome event itself leads to (sometimes extreme) difficulties with ensuing probability calculations, for no particular reason other than that we have decided to switch our choice of reference class from the infinite one defined by the model, where the relevant probability is intermediate and well defined, to the singleton one defined by the realized outcome, where the probability of the event was always only ever going to be either $0$ or $1$, regardless of when we assigned it.

More straightforwardly, and perhaps a bit provocatively, I would like to suggest that as a matter of mathematical bookkeeping, we might reevaluate the standard frequentist notion of probability as living only in the process of random sampling. The ``pre–data'' and ``post–data'' distinction already has an epistemic flavor, being naturally defined by someone’s having observed sampling to have taken place (otherwise, how could we tell between the two?), and it suggests that probability under the given model has somehow vanished once sampling has occurred. That implication is not required by frequentism per se, at least as realized in the Kolmogorov axioms: a Bernoulli–distributed outcome $Y_i$ determined by data $X_i$ carries an unconditional probability of success $p$ under the model, regardless of whether it has yet been realized. Instead, it seems more faithful to the frequentist program to ground our probability statements purely in the $\sigma$–algebras given to us by the model, choosing among them as needed so as to make our resulting inferences as accurate and as coherent as possible, and to resist the temptation to identify probability entirely with whatever ``chanciness'' may be inherent in physical, real–world processes, including a statistician drawing random samples from an actual population. I would argue that the latter picture is more properly viewed as a kind of frequentism–propensity–theory hybrid \cite{hajek2002interpretations, popper1959propensity}, and that we might not gain much in return by blurring those lines (if anything keeping them separate might help us see more clearly the pragmatic and philosophical benefits of both). On the other hand, hewing to a model-based view puts us directly in touch with the mathematical support we need to make precisely the sort of objective, long–run probability statements that frequentism was originally conceived to secure..

\section{5 Acknowledgments}

Thanks to Chad Heilig for helping me refine the mathematical ideas for this paper; to Mariette Marano-Lee for helping me refine the thought experiments; and to Cheri Gatland-Lightner for carefully reviewing the final draft.

\section{6 Conclusion}

In this paper, I have presented two kinds of arguments against the common claim that Neyman's interpretation of confidence intervals is the only one that may correctly be made: one informal, based on thought experiments, and one more formal, based on several views of the probability model governing coverage. Succinctly, the gist of both arguments is that the ``either-or'' logic Neyman appeals to, when taken seriously as a normative rule about when intermediate probability assignments are and are not allowed, causes substantial problems for probability calculations in frequentist settings. Moreover, that interpretation  sits uneasily with the very mathematical machinery Neyman uses to define the quantities he most cares about---namely, long-run error rates for the procedures he recommends. Those error rates are expectations of single-trial coverage indicators under the design, and thus presuppose non-degenerate, design-level probabilities for the corresponding single coverage events, precisely the sort of probabilities the strict ``either-or'' reading forbids us to use when interpreting a particular realized interval. 

As an alternative, or perhaps as a complement, I suggest that we may continue to report the design-level coverage probability ex post, deciding to view a constructed interval as simply one among many that could have been generated by the confidence procedure and not one with any particular numerical values attached to it. I also suggest that the notion of "confidence" refers very likely to the same mathematical concept as predictive probability, or as a probabilistic forecast, and that keeping that information-bound probabilistic layer separate from the two layers established by the underlying model can resolve some of the longstanding tensions around the use and interpretation of CIs for inference.

\printbibliography

\end{document}